\documentclass[preprint]{aastex62}
\usepackage{graphicx}
\usepackage{tablefootnote}

\newcommand{\magphys}{\hbox{\sc magphys}}
\newcommand{\cigale}{\hbox{\sc cigale}}
\newcommand{\bayesed}{\hbox{\sc bayesed}}
\newcommand{\clumpy}{\hbox{\sc clumpy}}
\newcommand{\wobs}{W0533$-$3401}
\newcommand{\wise}{{\textit{WISE~}}}

\newcommand{\herschel}{{\textit{Herschel~}}}

\newcommand{\coline}{CO$(3-2)$}

\hypersetup{linkcolor=red,citecolor=blue,filecolor=cyan,urlcolor=blue}

\shorttitle{ALMA reveals gas-rich maximum-starburst}
\shortauthors{Fan et al.}

\begin{document}

\title{ ALMA REVEALS A GAS-RICH, MAXIMUM-STARBURST IN THE HYPERLUMINOUS, DUST-OBSCURED QUASAR \wobs~ AT $z\sim2.9$}

\correspondingauthor{Lulu Fan}
\email{llfan@ustc.edu.cn}

\author[0000-0003-4200-4432]{Lulu Fan}

\affiliation{CAS Key Laboratory for Research in Galaxies and Cosmology, Department of Astronomy, University of Science and Technology of China, Hefei 230026, China}

\affil{School of Astronomy and Space Sciences, University of Science and Technology of China, Hefei, Anhui 230026, People's Republic of China}

\affil{Shandong Provincial Key Lab of Optical Astronomy and Solar-Terrestrial Environment, Institute of Space Science, Shandong University, Weihai, 264209, People's Republic of China}

\author[0000-0002-7821-8873]{Kirsten K. Knudsen}
\affiliation{Department of Space, Earth and Environment, Chalmers University of Technology, Onsala Space Observatory, SE-439 92 Onsala, Sweden}

\author[0000-0002-2547-0434]{Yunkun Han} 
\affiliation{Yunnan Observatories, Chinese Academy of Sciences, 396 Yangfangwang, Guandu District, Kunming, 650216, P. R. China}
\affiliation{Center for Astronomical Mega-Science, Chinese Academy of Sciences, 20A Datun Road, Chaoyang District, Beijing, 100012, P. R. China}
\affiliation{Key Laboratory for the Structure and Evolution of Celestial Objects, Chinese Academy of Sciences, 396 Yangfangwang, Guandu District, Kunming, 650216, P. R. China}

\author[0000-0003-3032-0948]{Qing-hua Tan} 
\affiliation{Purple Mountain Observatory \& Key Laboratory for Radio Astronomy, Chinese Academy of Sciences, 8 Yuanhua Road, Nanjing 210034, People's Republic of China}



\begin{abstract}

We present ALMA observations and multiwavelength spectral energy distribution (SED) analysis in a {\it WISE}-selected, hyperluminous dust-obscured quasar \wobs\ at $z=2.9$. We derive its physical properties of each component, such as molecular gas, stars, dust and the central supermassive black hole (SMBH). Both the dust continuum  at 3 mm and the \coline\ line are detected. The derived molecular gas mass $M_{\rm gas}=8.4\times10^{10}\ M_\odot$ and its fraction $f_{\rm gas}=0.7$ suggest that \wobs\ is gas-rich. The star formation rate (SFR) has been estimated to be $\sim3000-7000\ M_\odot$ yr$^{-1}$ by using different methods. The high values of SFR and specific SFR suggest that \wobs\ is a maximum-starburst. The corresponding gas depletion timescales are very short ($t_{\rm depl}\sim12-28$ Myr). The \coline\ emission line is marginally resolved and has a velocity gradient, which is possibly due to a rotating gas disk, gas outflow or merger.  Finally, we infer the black hole mass growth rate of \wobs\ (${\dot{M}}_{\rm BH}$ = 49 $M_\odot$ yr$^{-1}$), which suggests a rapid growth of the central SMBH. The observed black hole to stellar mass ratio $M_{\rm BH}/M_\star$ of \wobs, which is dependent on the adopted Eddington ratio, is over one order of magnitude higher than the local value, and is evolving towards the evolutionary trend of unobscured quasars. Our results are consistent with the scenario that \wobs, with both a gas-rich maximum-starburst and a rapid black hole growth, is experiencing a short transition phase towards an unobscured quasar. 

\end{abstract}



\section{Introduction} \label{sec:intro}

Super-massive black holes (SMBH) have been discovered in the centers of local elliptical galaxies, and the stellar bulge mass in galaxies is found to be correlated with the mass of their central SMBHs \citep[e.g.,][]{Magorrian1998,Ferrarese2005}, implying that galaxies and their central SMBHs could co-evolve \citep{Kormendy2013}. In the popular framework of massive galaxy formation and co-evolution with the central SMBH \citep[e.g.][]{Hopkins2008}, galaxy gas-rich mergers trigger intense starbursts and also provide the fuel for central SMBH accretion. An evolutionary sequence is predicted that the evolution of massive galaxies will experience several phases: starburst, dust-obscured quasar, unobscured quasar and finally a passively evolved galaxy \citep[e.g.,][]{Sanders1988,Granato2004,Alexander2012}. Dust-obscured quasars have been believed to represent a brief transition phase linking intense starbursts and unobscured quasars, and will be good candidates for studying the interplay between host galaxies and their central SMBHs\citep[e.g.,][]{Hickox2018}. 

Among several selection techniques commonly used to identify and characterize obscured quasar \citep[see a recent review by][]{Hickox2018}, the mid-IR color-color diagnostics \citep{Padovani2017} are efficient and effective for identifying heavily dust-obscured, powerful quasars. Recently, a new population of luminous, dust-obscured galaxies has been discovered based on a so-called {W1W2}-dropout color-selected method \citep{Eisenhardt2012,Wu2012} which uses only four mid-IR wavebands of {\it Wide-field Infrared Survey Explorer} \citep[{\it WISE};][]{Wright2010} all-sky survey.  Follow-up studies, including UV/optical spectral analysis \citep{Wu2012,Wu2018}, IR spectral energy distribution (SED) analysis \citep{Tsai2015,Fan2016b,Fan2018b}, X-ray observations \citep{Stern2014,Piconcelli2015,Assef2016,Ricci2017,Vito2018,Zappacosta2018} and high-resolution radio imaging \citep{Frey2016}, suggest that these {\it WISE}-selected galaxies are mainly powered by accreting SMBHs and substantially dust-obscured quasars. 

Multiwavelength observations have been carried out to investigate the physical properties of each component in these dust-obscured quasars, such as stars, dust, gas, the central SMBH, galaxy morphology and the environment they reside in \citep[e.g.,][]{Wu2014,Assef2015,Diaz-santos2016,Diaz-santos2018,Jones2015,Jones2017,Fan2016a,Fan2017a,Fan2018a, Tsai2018}. All results are generally consistent with the merger-driven SMBH-host co-evolution scenario. 


In this paper, we report the results of ALMA observations and a thorough UV-to-millimeter SED analysis of a {\it WISE}-selected dust-obscured quasar \wobs\ at $z\sim2.9$, which is among the most luminous obscured quasars with the total IR luminosity greater than $10^{14}L_\odot$ \citep{Tsai2015}. Our previous study based on IR SED decomposition suggests that \wobs\ has simultaneously rapid growth of SMBH and intense starburst ($>3000 M_\odot$ yr$^{-1}$) \citep{Fan2016b}. With the present ALMA observations and multiwavelength SED analysis, we can further derive the properties of each component in host galaxy and explore the potential relation between the assembly of host galaxy and the central SMBH accretion. In Section \ref{sec:alma}, we present our ALMA observations and data analysis on \wobs. In Section \ref{sec:multised}, we compile the multiwavelength data and introduce our SED modeling method. We show our main results and discussion in Section \ref{sec:res}. In Section \ref{sec:sum}, we summarize our conclusions. Throughout this work we assume a standard, flat ${\rm \Lambda}$CDM cosmology \citep[see][]{Komatsu2011}, with $H_0 = 70$ km~s$^{-1}$, $\Omega_M = 0.3$, and $\Omega_\Lambda = 0.7$.

\section{ALMA observations and data analysis} \label{sec:alma}

Observations of \wobs\ were obtained with ALMA using the band-3 receiver as a part of project 2017.1.00441.S. The observations were carried out on 2017-Dec-20 using 45 antennas in a configuration with baseline length ranging from 15m to 2460m. The on-source integration time was 517s. The sources J0538$-$4405 and J0522$-$3627 were used for bandpass and flux calibration, and for gain calibration, respectively.  The uncertainty on the absolute flux calibration was estimated to be about 5\%. The precipitable water vapour (PWV) was measured to be 4.2-4.5\,mm and the weather conditions were stable during this relatively short period. The  receiver settings were used as follows: the lower sideband has two spectral windows tuned to 87.418 and 89.219\,GHz with spectral line mode using 960 spectral channels each, where the tuning was selected to target \coline\ using the optical redshift $z_{\rm opt}=2.904$ \citep{Tsai2015}, and the upper sideband has two spectral windows tuned to 99.421 and 101.221\,GHz in continuum mode with 128 channels each.

The data were processed using CASA (Common Astronomy Software Application\footnote{\url{https://casa.nrao.edu/}}; \citealt{Mcmullin07}). We checked the data calibration from observatory delivered pipeline processing. We found that the calibration was sufficient and made no further adjustments. The calibrated visibilities were re-imaged using task {\tt tclean}. For natural weighting, the angular resolution of the observations (clean beam size) is $0.61''\times0.53''$, at a position angle (P.A.) = $-78.4$\,degrees, and the rms is 0.43\,mJy/beam in 53\,km\,s$^{-1}$ channels. A summary of target properties and ALMA measurements can be found in Table \ref{tab:obssum}.

\begin{deluxetable}{lc}
\tabletypesize{\scriptsize}
\tablecolumns{2}
\tablecaption{Summary of target properties and ALMA measurements\label{tab:obssum} }
\startdata
\\
Name                             &   \wobs          \\
R.A.$_{\it WISE}$ (J2000)        &   05:33:58.44     \\ 
Dec$_{\it WISE}$ (J2000)         &   $-34$:01:34.5   \\
$z_{\rm opt}$                    &   2.904           \\
\hline
Date of ALMA observations        &    2017-Dec-20               \\
Number of antennas               &    45                \\
R.A.$_{\rm CO(3-2)}$ (J2000)     &   05:33:58.42      \\ 
Dec$_{\rm CO(3-2)}$ (J2000)      &   $-34$:01:34.5    \\
$z_{\rm CO(3-2)}$               & $2.9026\pm0.0003$ \\
Size$_{\rm CO(3-2)}$ (arcsec$^2$)  &  (0.73 $\pm$ 0.14)$\times$(0.37 $\pm$ 0.14)                 \\
P.A.$_{\rm CO(3-2)}$ (deg)        &   126 $\pm$ 18        \\
FWHM (km~s$^{-1}$) \tablenotemark{a}  &   566 $\pm$ 44                    \\
$I_{\rm CO(3-2)}$ (Jy~km~s$^{-1}$)  & 2.01 $\pm$ 0.13                 \\
$L'_{\rm CO(3-2)}$ ($10^{10}$ K~km~s$^{-1}$~pc$^2$) & $8.4\pm0.5$    \\
$S_{\rm 3mm}$ [mJy]            &     0.140$\pm$0.033    \\
\enddata
\tablenotetext{a}{ From a single Gaussian fit.}
\end{deluxetable}

\section{Multiwavelength data and SED modeling} \label{sec:multised}

\subsection{UV-to-millimeter SED}\label{subsec:sed}

\begin{figure}
\plotone{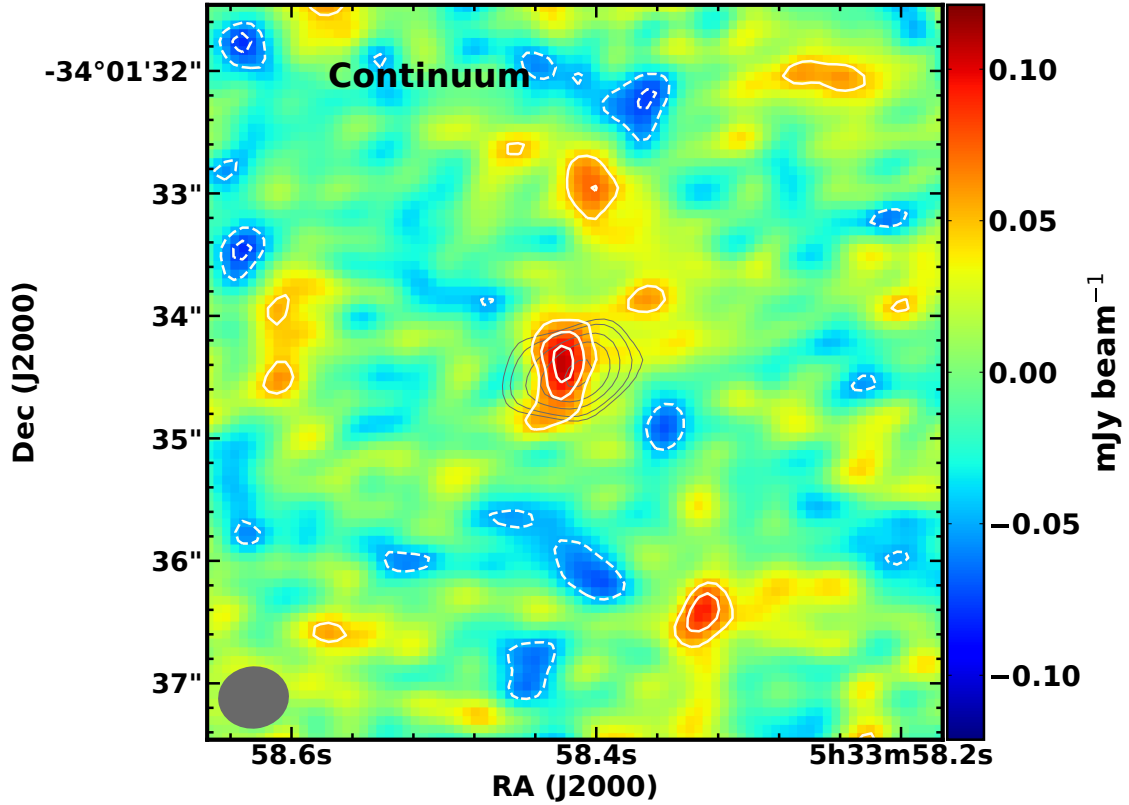}
\caption{ALMA 3mm dust continuum map of \wobs\ imaged with natural weighting. The ivory-colored contours are at $-3, -2, 2, 3$ and $4\sigma$ level. The beam size is shown in grey and is about $\sim0.6''\times0.5''$. The grey contours represent the natural weighted CO(3--2) moment-0 map at $4,5,7,9$ and $12\sigma$ level (see Fig.~\ref{fig:coimg01}).} \label{fig:dust}
\end{figure}

We construct the multiwavelength SED of \wobs\ by compiling the optical to millimeter broadband photometry from various catalogs available in the literature (see Table \ref{tab:photometry}). Optical/near-infrared photometry in five broad bands, $grizY$, are retrieved from the first public data release of the Dark Energy Survey \citep[DES DR1,][]{desdr1} \footnote{\url{https://des.ncsa.illinois.edu/releases/dr1/}}. Near-infrared $J$ band photometry has been obtained by SOAR/OSIRIS \citep{Assef2015}.  The \wise W3 and W4 photometry of \wobs\ are from the ALLWISE Data Release \citep{Cutri2013}. W3 and W4 flux densities and uncertainties have been converted from catalog Vega magnitude by using zero points of 29.04 and 8.284 Jy, respectively \citep{Wright2010}. While for \wise W1 and W2 photometry, we do the aperture photometry based on the unblurred coadded \wise images\footnote{\url{https://unwise.me/}} \citep[unWISE,][]{Lang2014,Meisner2017}. The photometry errors have been estimated based on the inverse variance images. We also collect the FIR photometry of \wobs\ obtained with \herschel \citep{Pilbratt2010} PACS \citep{Poglitsch2010}  at 70 and 160 $\mu$m and SPIRE \citep{Griffin2010} at 250, 350 and 500 $\mu$m in our previous work \citep{Fan2016b}. Our ALMA observations show a marginal detection ($\sim4.2\sigma$) of 3\,mm dust continuum. The measured observed-frame continuum flux density is $0.140\pm0.033$ mJy (see Figure \ref{fig:dust}). 

\subsection{Multiwavelength SED analysis}\label{subsec:sed}

\begin{figure*}
\plotone{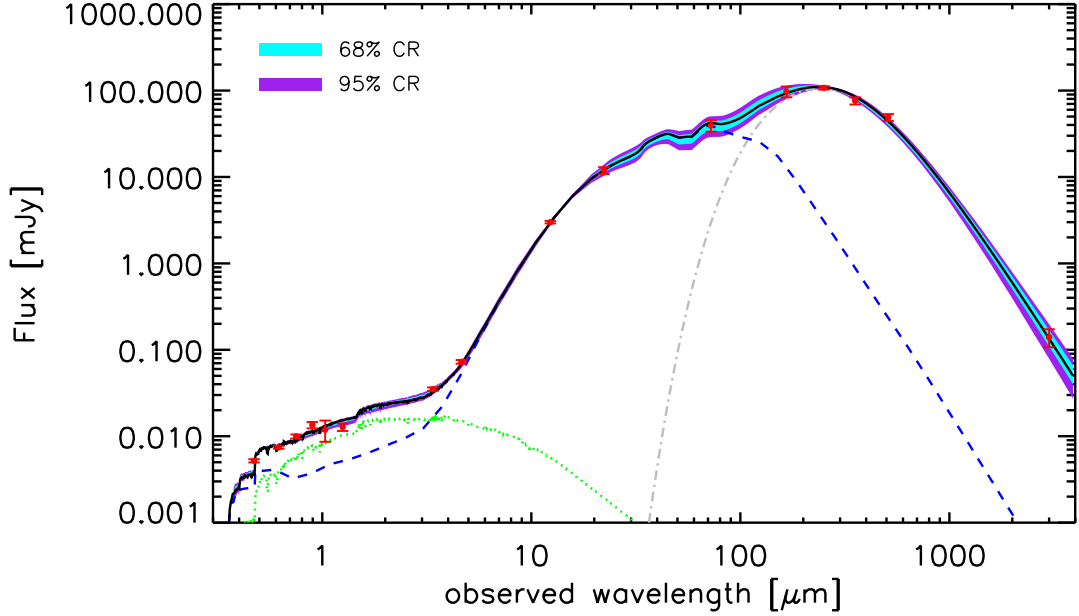}
\caption{The best-fit three-component SED model (black solid line) with \bayesed\ to the observed UV-to-millimeter SED of \wobs\ (red points). The color-filled regions with cyan and purple show the 68\% and 95\% CR of the best-fit model, respectively. Green dotted line, blue dashed line and gray dot-dashed line represent the emissions from stars, AGN and cold dust, respectively. \label{fig:sedfitting}}
\end{figure*}

\begin{deluxetable*}{lccc}
\centering
\tabletypesize{\small}
\tablecolumns{4}
\tablecaption{Multiwavelength photometry data \label{tab:photometry}}
\tablehead{
\colhead{Band} & \colhead{Wavelength}  & \colhead{Frequency} & \colhead{Flux Density} \\
\colhead{ }    & \colhead{($\micron$)} & \colhead{(GHz)}     & \colhead{(mJy)}
}
\startdata
CTIO/DECam $g$            &  0.36     & 833333  & 0.0052 $\pm$ 0.0002  \\
CTIO/DECam $r$            &  0.54     & 555556  & 0.0075 $\pm$ 0.0003  \\
CTIO/DECam $i$            &  0.64     & 468750  & 0.0099 $\pm$ 0.0005  \\
CTIO/DECam $z$            &  0.77     & 389610  & 0.0135 $\pm$ 0.0011  \\
CTIO/DECam $Y$            &  0.90     & 333333  & 0.0119 $\pm$ 0.0033  \\
SOAR/OSIRIS $J$           &  1.05     & 285714  & 0.0128 $\pm$ 0.0013  \\
{\it WISE}/W1           &   3.4     & 88174.2 & 0.0351 $\pm$ 0.0016     \\
{\it WISE}/W2           &   4.6     & 65172.3 & 0.0726 $\pm$ 0.0036    \\
{\it WISE}/W3           &   12      & 24982.7 & 3.0    $\pm$ 0.1    \\
{\it WISE}/W4           &   22      & 13626.9 & 11.9   $\pm$ 1.1    \\
{\it Herschel}/PACS     &   70      & 4143.65 & 39.3   $\pm$ 5.9      \\
{\it Herschel}/PACS     &   160     & 1805.05 & 97.4   $\pm$ 14.0      \\
{\it Herschel}/SPIRE    &   250     & 1199.17 & 107.5  $\pm$ 4.8      \\
{\it Herschel}/SPIRE    &   350     & 856.55  & 76.3   $\pm$ 7.3      \\
{\it Herschel}/SPIRE    &   500     & 599.585 & 48.9   $\pm$ 4.5    \\
ALMA                    &   3000    & 100.00  & $0.140\pm0.033$   \\
\enddata
\end{deluxetable*}

For the multiwavelength SED analysis of \wobs, we use a forthcoming version (Han et al., in prep) of the Bayesian SED modeling and interpreting code \bayesed \footnote{\url{https://bitbucket.org/hanyk/bayesed/}}\citep{Han2012a,HanY2014a,HanY2019a}. In the new version of \bayesed, the stellar emission, dust attenuation, and dust emission can be consistently connected by assuming an energy balance, a technique similar to that employed in \magphys\ \citep{dacunha2008} and \cigale\ \citep{Noll2009a,BoquienM2019a}. The stellar emission is modeled by using the \cite{bc03} SSP with a \cite{chabrier2003} initial mass function (IMF), an exponentially declining star formation history (SFH) and the \cite{Calzetti2000a} dust attenuation law. The energy of stellar emission absorbed by dust is assumed to be totally re-emitted at IR band, which is modeled by a graybody. The graybody model is defined as: $S_{\lambda}\propto(1-e^{-(\frac{\lambda_0}{\lambda})^{\beta}}) B_\lambda(T_{dust})$,  where $B_\lambda$ is the Planck blackbody spectrum and $\lambda_0$ = 125$\,\mu$m. Dust temperature $T_{\rm dust}$ and the emissivity index $\beta$ are two free parameters. Finally, as in \cite{Fan2016b}, the AGN torus emission is modeled independently with the extensive database \footnote{\url{http://www.pa.uky.edu/clumpy/models/ clumpy_models_201410_tvavg.hdf5/}} of 1,247,400 SEDs from the \clumpy\ torus model \citep{Nenkova2008a,Nenkova2008b}. A k-dimensional tree based nearest-neighbor searching technique has been employed to allow us to evaluate the \clumpy\ torus model at any point in its 6-D parameter space. We remind that the \clumpy\ model includes not only the torus dust emission, but also a part of the AGN accretion disk emission that is scattered into our line of sight or not absorbed by torus dust. Thus the \clumpy\ model can provide a consistent description of AGN UV-to-millimeter SED. In total, the three-component model, including stars, AGN and cold dust emissions,  has 12 free parameters. The priors for them are summarized in Table \ref{tab:priors}.

In Figure \ref{fig:sedfitting}, we show the best-fit three-component SED model (black solid line) with \bayesed\ to the observed UV-to-millimeter SED of \wobs\ (red points). With a newly-developed function in the new version of \bayesed, we can provide the confidence regions (CR) for our best-fit SED model. We plot the 68\% and 95\% CR with color-filled regions (cyan and purple, respectively). Absorbed stellar emission, AGN emission and cold dust emission have been shown in green dotted line, blue dashed line and gray dot-dashed line, respectively. The derived properties have been shown in Table \ref{tab:obsprop} and the results will be discussed in the next section.

\begin{deluxetable*}{lllll}
\tablecolumns{5}
\tabletypesize{\small}
\tablewidth{0pt}
\tablecaption{The model parameters, their priors and best-fitting quantities for multiwavelength SED analysis \label{tab:priors}}
\tablehead{
    \colhead{Name} & \colhead{Prior} & \colhead{Min} & \colhead{Max} & \colhead{Best-fitting value}
}
\startdata
        \multicolumn{5}{c}{Stellar component $^{a}$} \\
		\hline
		${\rm log}(age/{\rm yr})$ & Uniform, $age<age_{\rm U}(z)$ & 5 & 10.3 & $6.7^{+0.1}_{-0.1}$ \\
		${\rm log}(\tau/{\rm yr})$ & Uniform & 6 & 12 & $9.1^{+1.8}_{-1.9}$ \\
		${\rm log}(Z/{\rm Z_{\odot}})$ & Uniform & -2.3 & 0.7 & $0.28^{+0.18}_{-0.38}$ \\
		$A_{\rm V}/{\rm mag}$ & Uniform & 0 & 4 & $1.81^{+0.08}_{-0.08}$ \\
		\hline
		\multicolumn{5}{c}{Graybody model} \\
		\hline
		$T_{\rm dust}/{\rm K}$ & Uniform & 10 & 100 & $78.1^{+5.9}_{-5.2}$ \\
		$\beta$ & Uniform & 1 & 3 & $1.84^{+0.14}_{-0.13}$ \\
		\hline
		\multicolumn{5}{c}{\clumpy\ model $^{b}$} \\
		\hline
		$N_0$ & Uniform & 1 & 15 & $5.9^{+1.3}_{-1.0}$ \\
		$Y$ & Uniform & 5 & 100 & $48.2^{+32.6}_{-28.2}$ \\
		$i$ & Uniform & 0 & 90 & $37.8^{+20.0}_{-23.8}$ \\
		$q$ & Uniform & 0 & 3 & $2.2^{+0.4}_{-0.4}$ \\
		$\sigma$ & Uniform & 15 & 70 & $62.3^{+8.5}_{-5.1}$ \\
		$\tau_{\rm V}$ & Uniform & 10 & 300 & $22.2^{+5.4}_{-5.1}$ \\
\enddata
\tablecomments{$^{a}$ Four parameters of stellar components: age ($age$), the e-folding time for the exponentially declining SFH ($\tau$), stellar metallicity ($Z$) and dust attenuation ($A_{\rm V}$). 
$^{b}$ The detailed description and definition of six free parameters for \clumpy\ model can be found in \cite{Nenkova2008a,Nenkova2008b}.}
\end{deluxetable*}

\section{Results and discussion} \label{sec:res} 

\subsection{CO(3-2) line emission}\label{subsec:gas}

\begin{figure*}
\includegraphics[width=0.49\textwidth]{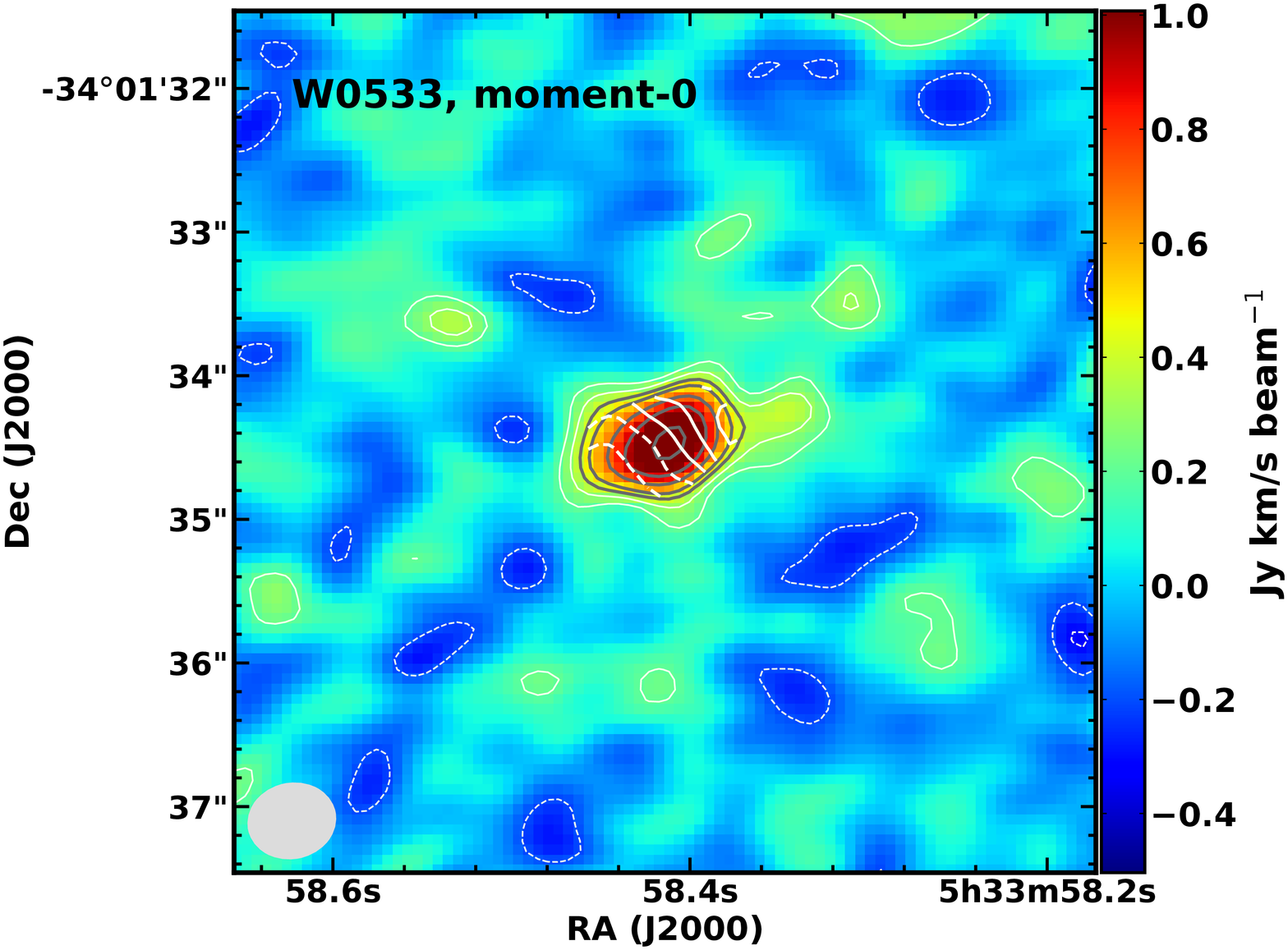}
\includegraphics[width=0.49\textwidth]{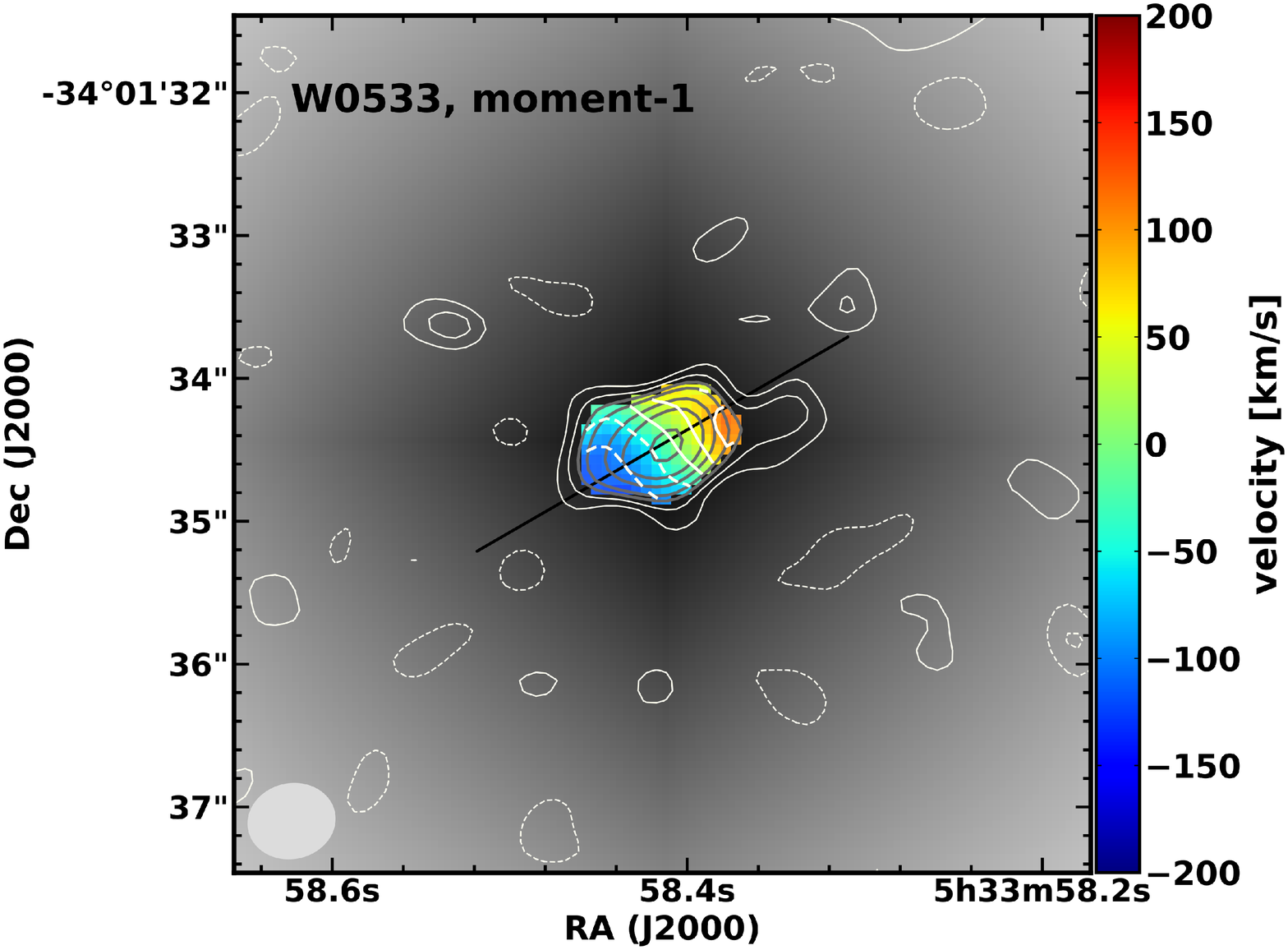}
\caption{In the left panel, we plot moment-0 map of the \coline\ emission line, where the grey contours represent the moment-0 at $4,5,7,9,12\sigma$ and the ivory contours are at $-3, -2, 2, 3\sigma$ level ($\sigma \sim 0.1$\,Jy\,km/s\,beam$^{-1}$). In the right panel, the ivory contours represent the moment-1 in steps of 50\,km\,s$^{-1}$. The black line shows the direction of the major axis. The beam size is shown in the bottom-left corner of each plot.}
\label{fig:coimg01}
\end{figure*}

\begin{figure}
\plotone{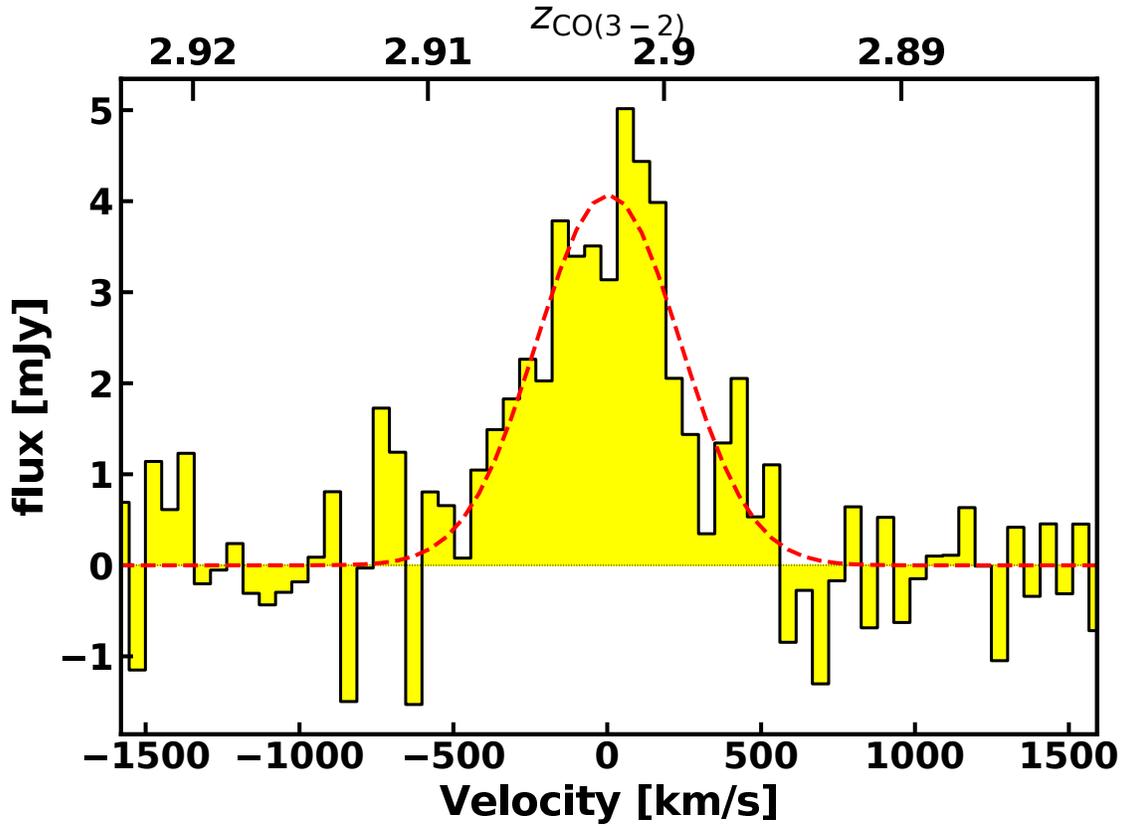}
\caption{The continuum-subtracted \coline\ spectrum. The red dashed line shows
a fit with a Gaussian function.\label{fig:cospec}}
\end{figure}

In Figures~\ref{fig:coimg01} and \ref{fig:cospec}, we show the resulting line detection, both as spectrum and moment-maps. The \coline\ emission line peaks at a frequency corresponding to a redshift of $z_{\rm CO(3-2)}=2.9026\pm0.0003$. The detected \coline\ line is spatially extended and shows a velocity gradient. We fit the velocity-integrated visibilities with an elliptical Gaussian function. The derived source size deconvolved from beam is (0.73 $\pm$ 0.14)$\times$(0.37 $\pm$ 0.14) arcsec$^2$ with P.A. = $126\pm18$. We measure the line flux $I_{\rm CO(3-2)}=2.01\pm0.13$ Jy km s$^{-1}$. 
The rms is determined in the inner 10$''$ of the moment map excluding the central part that contains the source itself.
A single Gaussian fit to the continuum-subtracted \coline\ spectrum gives \coline\ FWHM = $566\pm44$ km s$^{-1}$. From the \coline\ line flux, we can derive the CO line luminosity $L'_{\rm CO(3-2)}=(8.4\pm0.5)\times10^{10}$ K~km~s$^{-1}$~pc$^2$, by using Equation 3 in \citet{solomon2005}. The continuum is marginally detected, as shown in Figure~\ref{fig:dust}. The extend of the continuum emission is smaller than that of the \coline\ line, which is likely due to a combination of the optical depth effects and the signal-to-noise ratio of continuum detection. The results are summarized in Table~\ref{tab:obssum}.

We can infer the molecular gas mass directly from the measured CO line luminosity. We adopt the excitation ratio between the \coline\ and CO$(1-0)$ line of $r_{32/10} = 0.8$ as suggested by \cite{Banerji2017}. The adopted ratio is intermediate between the typical values for submillimeter galaxies (SMGs, $r_{32/10} = 0.66$) and optical quasars ($r_{32/10} = 0.97$) from \cite{Carilli2013}, and consistent with the expectation that obscured quasars represent the transition phase from SMGs to unobscured quasars. The calculated CO$(1-0)$ line luminosity is $L'_{\rm CO(1-0)}=(10.5\pm0.6)\times10^{10}$ K~km~s$^{-1}$~pc$^2$. We also adopt the CO-to-H$_2$ conversion factor, $\alpha_{\rm CO}=0.8~M_\odot$(K km s$^{-1}$ pc$^2$)$^{-1}$, which is suggested to be appropriate for starbursts and quasar hosts \citep{Carilli2013}. Under these considerations, we obtain a molecular gas mass of $M_{\rm H_2}=(8.4\pm0.5)\times10^{10}\,M_\odot$. 

\begin{figure}
\plotone{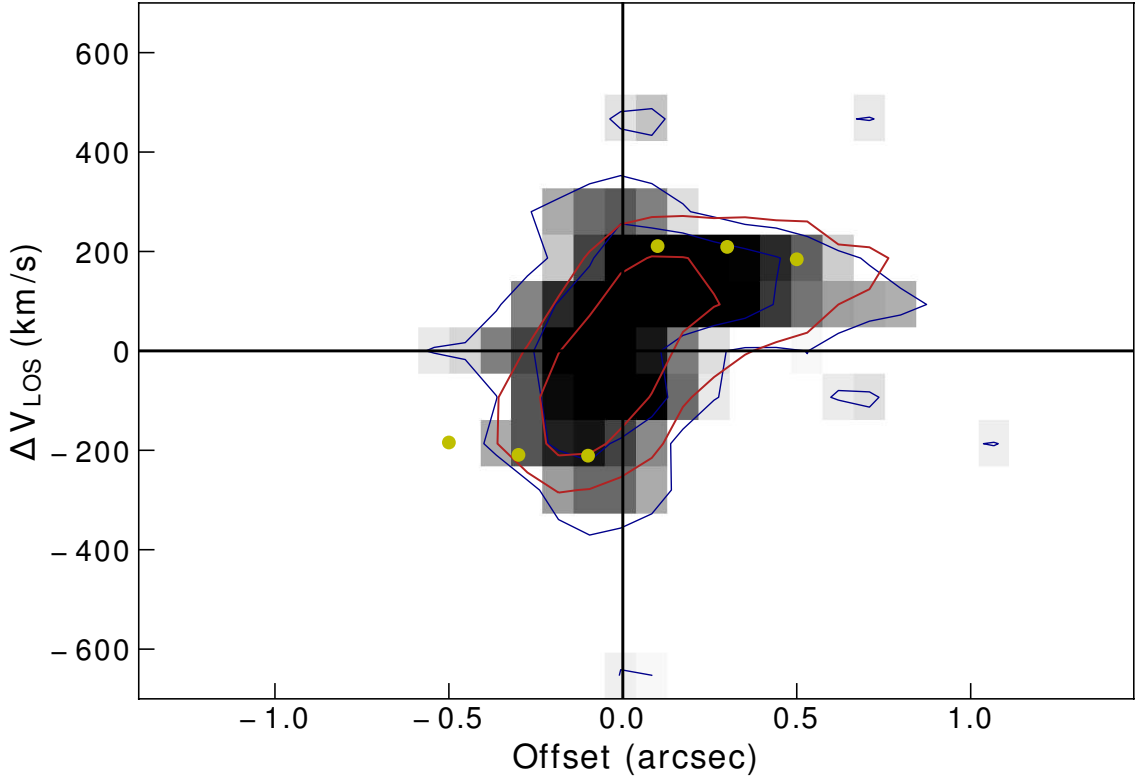}
\caption{\coline\ Position versus velocity (PV) diagram extracted along the major axis at P.A.=330$^{\circ}$ assuming an inclination of 59$^{\circ}$. The blue and red contours indicate the iso-density contours of the galaxy and  {\tt $^{3D}$BAROLO} best-fit model. The yellow points mark the rotation curve with rotation velocities of 246, 244, and 215 km s$^{-1}$ (from center to outer disk), assuming a disc model with three rings and each ring width of 0.2$''$. \label{fig:pv}}
\end{figure}

We show the \coline\ velocity map in the right panel of Figure \ref{fig:coimg01} and the Position versus velocity (PV) diagram extracted along the major axis at P.A. = 330$^{\circ}$ in Figure \ref{fig:pv}. The \coline\ emission line is marginally resolved. Both figures show the possible presence of a velocity gradient. Such a velocity gradient, which has also been observed in many other high-redshift quasars \citep[e.g.,][]{Leung2017,Brusa2018,Feruglio2018,Talia2018},  could be possibly the signature of a rotating disk of molecular gas, although the explanation is not unique. Especially, tentatively emission is seen at $\sim3\sigma$ level extending to the west side of the source (see the left panel of Figure \ref{fig:coimg01}). This suggests that the gas-rich late-stage major merger and gas outflow are also possible for originating such a velocity gradient \citep[e.g.,][]{Springel2005,Hopkins2009,Ueda2014,Hung2015}. However, due to limited depth and resolution, we cannot make a solid conclusion. Deeper observation would be required to distinguish them.

Assuming that the gas is distributed in a rotating disk, we can investigate the kinematic properties of the molecular gas traced by the \coline\ line, using {\tt $^{3D}$BAROLO}, a tool for fitting 3D tilted-ring models to emission-line data cubes \citep{DiTeodoro2015}. We assume a disc model with three rings and a ring width of 0.2$''$. We fix P.A.$ = 330^{\circ}$ and adopt an inclination of  59$^{\circ}\pm 14^{\circ}$, which is inferred from the observed ratio of minor to major axis. The derived rotation velocity ($V_{\rm rot}$) and the intrinsic velocity dispersion $\sigma$ are about $\sim240$ km s$^{-1}$ and 60 km s$^{-1}$, respectively. The ratio of rotation velocity to velocity dispersion $V_{\rm rot}/\sigma$ for \wobs\ is about 4, which is close to the typical value $\sim7$ for the molecular gas in $z>1$ star-forming galaxies \citep{Tacconi2013}. The large ratio $V_{\rm rot}/\sigma$ indicates that the molecular gas is turbulent, which is possibly due to a thick, dynamically hot disk in \wobs. A similar finding has been reported by \cite{Tadaki2018} that they found a gravitationally unstable, rotating gas disk in an extreme starburst galaxy at $z\sim4$. The dynamical mass within the CO-emitting region can be estimated by applying the relation in \cite{Wang2013}: $M_{\rm dyn}/M_\odot=1.16\times10^5\times(0.75\times{\rm FWHM_{\rm CO}})^2\times D/{\rm sin}i$, where $D$ is the disk diameter in kpc from the \coline\ measurement and $i$ is the inclination angle. This gives $M_{\rm dyn}=1.6\times10^{11}M_\odot$.

\subsection{Dust properties}\label{subsec:dust}

Cold dust emission heated by stars has been modeled with a graybody model in our  multiwavelength SED analysis (see Section \ref{subsec:sed} and Figure \ref{fig:sedfitting}). The best-fit model gives the estimations of dust properties: IR luminosity $L_{\rm GB}$, dust temperature $T_{\rm dust}$ and the emissivity index $\beta$, which are listed in Table \ref{tab:obsprop}. The derived values are well consistent with those in our previous work based on only IR SED decomposition \citep{Fan2016b}. Dust mass can also be calculated by using the following equation:
\begin{equation}\label{equ:mdust}
    M_{\rm dust}=\frac{D^2_{\rm L}}{(1+z)}\times\frac{S_{\nu_{\rm obs}}}{\kappa_{\nu_{\rm rest}} B(\nu_{\rm rest},T_{\rm dust})}
\end{equation}
where $D_{\rm L}$ is the luminosity distance, $S_{\nu_{\rm obs}}$ is the flux density at observed frequency $\nu_{\rm obs}$, $\kappa_{\nu_{\rm rest}}=\kappa_0(\nu/\nu_0)^\beta$ is the dust mass absorption coefficient at the rest frequency of the observed band, and $B(\nu_{\rm rest},T_{\rm dust})$ is the Planck function at temperature $T_{\rm dust}$. Adopting the best-fit values  $T_{\rm dust}=78.1$ K, $\beta=$1.84 and $\kappa_{\rm 1THz}=20$ cm$^2$ g$^{-1}$ which is the same as in \cite{Wu2014} and \cite{Fan2016b}, we derive the dust mass $M_{\rm dust}=8.9\pm0.5\times10^7M_\odot$. The small uncertainty of dust mass estimation only takes the uncertainties of the derived  $T_{\rm dust}$ and $\beta$ values into account. We note that the largest uncertainty can arise from the adopted $\kappa_{\nu_{\rm rest}}$ value, which can vary by over one order of magnitude at a certain frequency/wavelength. For instance, $\kappa_{850\mu m}$ can vary from $\sim$0.4 to $\sim$11 cm$^2$ g$^{-1}$ in the literature \citep[e.g.,][]{james2002,dunne2003,draine2003,Siebenmorgen2014}. 

We derive the gas-to-dust mass ratio of \wobs, $\delta_{\rm GDR} = 944\pm77$, based on the estimations of $M_{\rm H_2}$ and $M_{\rm dust}$. The derived $\delta_{\rm GDR}$ value of \wobs\ is thus about one order of magnitude higher than the typical value $\sim50-150$ derived for the Milky Way \citep{Jenkins2004}, the local star-forming galaxies (SFGs) and ultraluminous IR galaxies (ULIRGs) at solar metallicity \citep{Draine2007,Remy2014} and high-redshift SMGs \citep{Magnelli2012,Miettinen2017}. The high $\delta_{\rm GDR}$ value in \wobs\ may be due to several possible reasons: the uncertainty of dust mass estimation, the low efficiency of dust formation and/or the high efficiency of dust destruction. The dust mass derived by using a graybody model is about half of that derived by \cite{Draine2007b} model \citep{Magdis2011}. This will result in an overestimation of $\delta_{\rm GDR}$ by a factor of two. The $\delta_{\rm GDR}$ value is reported to increase with the decreasing metallicity and the increasing redshift \citep[e.g.,][]{Remy2014,Miettinen2017}. It is possible that \wobs\ has a low, sub-solar metallicity. It is also possible that dust destruction in \wobs\ may be efficient due to the strong radiation field from massive young stars and AGN, and the supernova shock waves which are expected to be frequent in this maximum-starburst galaxy \citep{Jones2004}. 

    
\subsection{Stellar mass and star formation rate}\label{subsec:stars}

\begin{deluxetable}{lclc}
\tabletypesize{\small}
\tablecolumns{4}
\tablecaption{Derived Properties of \wobs \label{tab:obsprop}}
\tablehead{ \colhead{Parameter} & \colhead{Value}     & \colhead{Unit}    & \colhead{Note}  }
\startdata
$M_\star$                    &  (3.5 $\pm$ 0.9) $\times10^{10}$  &   $M_\odot$   & (1)    \\ 
SFR                          &  6985 $\pm$ 3006         &   $M_\odot$ yr$^{-1}$    & (2) \\ 
${\rm sSFR}$                 &  200 $\pm$ 100           &      Gyr$^{-1}$    & (3)     \\
$L_{\rm AGN}$                &  (7.0 $\pm$ 0.6)$\times10^{13}$            &   $L_\odot$     & (4) \\ 
$L_{\rm GB}$                 &  (3.5 $\pm$ 0.4)$\times10^{13}$             &   $L_\odot$    & (5) \\ 
$L_\star^{\rm unabs}$        &  (3.6 $\pm$ 0.4)$\times10^{13}$             &    $L_\odot$   & (6) \\ 
$L_\star^{\rm abs}$          &  (9.7 $\pm$ 1.0)$\times10^{11}$             &    $L_\odot$           & (7) \\ 
$M_{\rm H_2}$                &  ($8.4\pm0.5$)$\times10^{10}$       &    $M_\odot$        & (8) \\ 
$M_{\rm dyn}$                &  $1.6\times10^{11}$              &  $M_\odot$        & (9) \\ 
$M_{\rm BH}$                 &  $2.2\times10^9$           &       $M_\odot$        & (10) \\ 
$M_{\rm dust}$               &  (8.9 $\pm$ 1.6) $\times 10^{7}$  &     $M_\odot$         & (11) \\ 
T$_{\rm dust}$               &  78.1 $\pm$ 5.6                &   K         & (12) \\ 
$\beta$                      &  1.84 $\pm$ 0.14               &              & (13) \\
${\rm A_V}$                  &  1.81 $\pm$ 0.08               &   mag      & (14) \\
${\rm t_{depl}}$             &  $\sim12-28$                 &   Myr        & (15)   \\
\enddata
\tablecomments{(1): Stellar mass; (2): Star formation rate; (3): Specific star formation rate; (4): Bolometric luminosity of AGN component; (5): IR luminosity of cold dust; (6): Bolometric luminosity of attenuation-corrected stellar emission; (7): Bolometric luminosity of absorbed stellar emission; (8): Molecular gas mass; (9): Dynamical mass; (10): The central suppermassive black hole mass by assuming $\lambda_{\rm Edd}=1$; (11): Dust mass; (12): Dust temperature; (13): Dust emissivity index; (14): Dust attenuation; (15): The gas depletion timescales
}
\end{deluxetable}

\begin{figure*}
\plotone{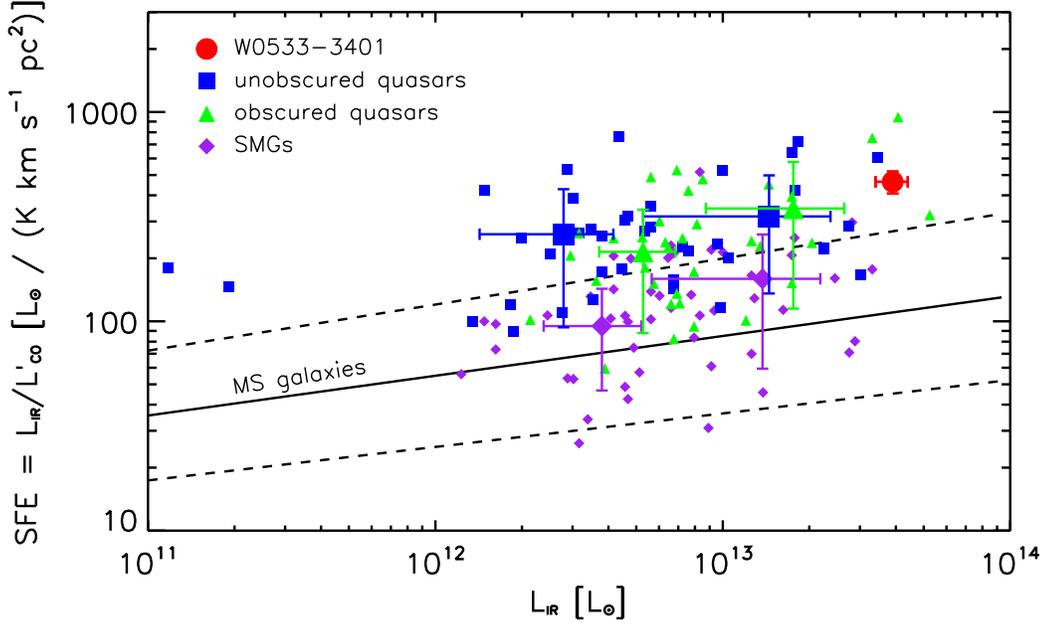}
\caption{Star formation efficiency (SFE) traced by the ratio of IR to CO$(1-0)$ line luminosities as a function of IR luminosity. \wobs\ is labelled with a red circle symbol. The purple diamond, blue square and green triangle symbols represent the samples of SMGs, unobscured and obscured quasars at $z>1$, respectively, compiled in \cite{Perna2018}. Each sample has been divided into two bins according to the IR luminosity. Larger symbols with error bars show the average values and uncertainties for each bin. The solid line and two dashed lines show the best-fit relation and 1$\sigma$ scatter for massive MS galaxies, respectively \citep{Sargent2014}.  \label{fig:sfe}}
\end{figure*}

Based on our best-fit SED model presented in Section \ref{subsec:sed}, we derive the stellar mass and star formation rate (SFR) of \wobs\ and list them in Table \ref{tab:obsprop}. The stellar mass $M_\star$ is derived by adopting an exponentially declining SFH, which is represented by a young stellar population. We check if there is an old stellar population which is omitted by the present SED fitting procedure. We add an SSP with an age of 1\,Gyr to our multiwavelength SED model. The result is that the contribution of the old SSP to the derived stellar mass is not larger than 20\%, though the constraint from the SED fitting is loose. Given $M_\star=3.5\times10^{10}M_\odot$ and $M_{\rm H_2}=8.4\times10^{10}M_\odot$, we can calculate the molecular gas fraction $f_{\rm gas}=M_{\rm H_2}/(M_{\rm H_2}+M_\star)=0.71$, which indicates that \wobs\ is gas-rich. The derived SFR is $\sim7000\ M_\odot$yr$^{-1}$.  The specific SFR or sSFR = 200 Gyr$^{-1}$ of \wobs\ is over one order of magnitude higher than the star-forming main-sequence (MS) at $z\sim3$ \citep{Speagle2014}, suggesting that it is a maximum-starburst. The uncertainty of SFR is large ($\sim3000\ M_\odot$yr$^{-1}$) due to a wide range of possible SFH. By using CO line luminosity, we can estimate SFR with an individual way. Firstly, we convert the measured $L'_{\rm CO(3-2)}$ to $L'_{\rm CO(5-4)}$ by taking $L'_{\rm CO(5-4)}/L'_{\rm CO(3-2)}=0.7$ which is intermediate between the typical values for SMGs and quasars \citep{Carilli2013}. Then we estimate FIR luminosity using high-J CO versus FIR luminosity relation presented in \cite{Liu2015}. Using the relation SFR ($M_\odot$yr$^{-1}$)=$4.5\times10^{-44}L_{\rm FIR}$ (erg s$^{-1}$), we calculate SFR$\sim3000\ M_\odot$yr$^{-1}$ \citep{Kennicutt1998}, which is generally consistent with the best-fit SED result. From the derived SFR and molecular gas mass, we can estimate the gas depletion timescale $t_{\rm depl}=M_{\rm gas}/$SFR. We infer $t_{\rm depl}\sim 12-28$ Myr using SFRs based on the best-fit SED result and the CO line luminosity, respectively. The gas depletion timescale of \wobs\ is similar to other obscured quasars \citep[e.g.,][]{Aravena2008,Brusa2018}, but is much shorter than MS galaxies and SMGs \citep{Bothwell2013,Sargent2014}, indicating that \wobs\ as an obscured quasar is consuming its residual gas more rapidly.

In Figure \ref{fig:sfe}, we report the star formation efficiency (SFE), traced by the ratio of IR to CO$(1-0)$ line luminosities, as a function of IR luminosity for \wobs\ and the compiled samples of SMGs, unobscured and obscured quasars at $z>1$ in \cite{Perna2018}. For all three samples, SFE shows a positive correlation with IR luminosity.  \wobs\ shows a high SFE=$L_{\rm IR}/L'_{\rm CO}=464\pm56\ L_\odot/$(K km s$^{-1}$ pc$^2$), which is well above the best-fit relation and 1$\sigma$ scatter for massive MS galaxies \citep{Sargent2014}. Similar as \wobs, both unobscured and obscured quasars at $z>1$ show a higher SFE at a given IR luminosity than MS galaxies and SMGs. Higher SFE in unobscured and obscured quasars relative to MS galaxies and SMGs is possibly due to the presence of starburst activity and/or the depletion of cold gas by AGN feedback in quasar host galaxies. The former enhances $L_{\rm IR}$ and the latter reduces $L'_{\rm CO}$. It is possible that $L_{\rm IR}$ can be overestimated in quasars, as AGN-heated dust on kpc scales can contribute significantly to IR luminosity \citep{Duras2017,Symeonidis2017}. 

\begin{figure*}
\plotone{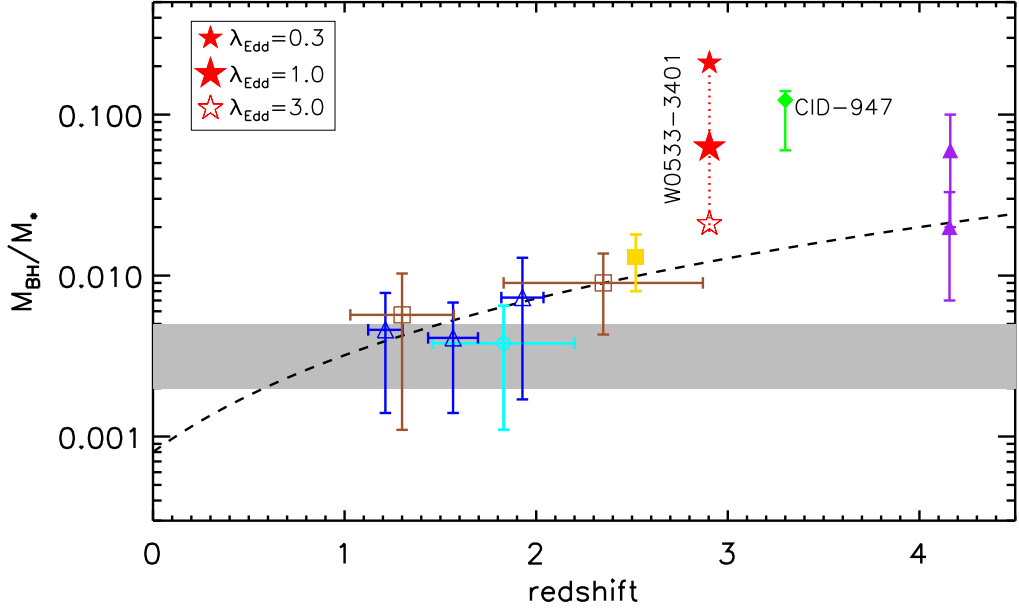}
\caption{The observed cosmic evolution of black hole to stellar mass ratio ($M_{\rm BH}/M_\star$). The $M_{\rm BH}/M_\star$ ratios of \wobs\ are shown as three star symbols by assuming different Eddington ratios $\lambda_{\rm Edd}=L_{\rm bol}/L_{\rm Edd}=$ 0.3, 1.0 and 3.0, respectively. Other data points represent the available $M_{\rm BH}/M_\star$ estimates in the literature. Color-coded symbols show the average values and the corresponding uncertainties of $M_{\rm BH}/M_\star$ for several samples: blue triangles for a sample of broad-line AGN in the redshift range $1<z<2.2$ \citep{Merloni2010}, brown squares for a sample of lensed and nonlensed quasars at $1<z<4.5$ \citep{Peng2006}, and cyan diamond for a sample of X-ray obscured, red AGN \citep{Bongiorno2014}. Two luminous SDSS quasars at $z\sim4$ are denoted with purple filled triangles \citep{Targett2012}. Gold filled square shows an extremely red dust-obscured quasar (WISE J1042+1641) at $z = 2.52$ \citep{Toba2016,Matsuoka2018}. Green filled diamond denotes an X-ray selected unobscured quasar, CID-947 at $z\sim3.3$, which has a high black hole to stellar mass ratio $M_{\rm BH}/M_\star=1/8$ \citep{Trakhtenbrot2015}. The dashed line shows the evolutionary trend of $M_{\rm BH}/M_\star$ at $z<2$ \citep{McLure2006}. The gray area shows the range of the typical values of $M_{\rm BH}/M_\star$  in the local Universe \citep{Kormendy2013}.  \label{fig:mbh2mstar}}
\end{figure*}

\subsection{Rapid growth of both the stellar component and the central SMBH}

We derive the AGN bolometric luminosity $L_{\rm AGN}$ of \wobs\ by integrating the AGN component of the best-fit UV-to-millimeter SED (see Section \ref{subsec:sed}). Although we do not have a direct measurement of the black hole mass in \wobs, we can make a rough estimate from the derived $L_{\rm AGN}$. By assuming Eddington ratios of 0.3, 1.0 and 3.0, the corresponding black hole masses are $7.3\times10^9$, $2.2\times10^9$ and $7.3\times10^8\ M_\odot$, respectively.  \cite{Wu2018} reported the black hole masses and Eddington ratios of five luminous obscured quasars at $z\sim2$, which are taken from the same parent sample of \wobs, based on broad H$_\alpha$ lines. They found that the average black hole mass is about $10^9\ M_\odot$ and the derived Eddington ratios are close to unity. \cite{Tsai2018} also reported the measurement of $M_{\rm BH}$ and $L_{\rm Edd}$ for an extremely luminous, obscured quasar W2246$-$0526 at $z=4.6$, which is selected with the same criterion as \wobs. They found that the central SMBH in it is growing rapidly by accreting at a super-Eddington ratio ($\lambda_{\rm Edd}=2.8$). Adopting the  Eddington ratio $\lambda_{\rm Edd}=1.0$ for \wobs\ will be a reasonable approximation.

From SED-based stellar mass $M_\star$ and the estimated SMBH mass, we can infer the black hole to stellar mass ratio of \wobs. In Figure \ref{fig:mbh2mstar}, we plot the observed $M_{\rm BH}/M_\star$ as a function of redshift for \wobs\ and several other samples in the literature. If assuming $\lambda_{\rm Edd}=1.0$,  the black hole to stellar mass ratio of \wobs\ ($M_{\rm BH}/M_\star=0.063$) is close to that of an X-ray selected unobscured quasar, CID-947, which has the highest $M_{\rm BH}/M_\star$ (1/8) at $z\sim3.3$ known so far. A low Eddington ratio in \wobs, for instance $\lambda_{\rm Edd}<0.3$, seems not likely, otherwise $M_{\rm BH}/M_\star$ of \wobs\ will be higher than 0.2. Even if taking $\lambda_{\rm Edd}=1.0$, the inferred $M_{\rm BH}/M_\star$ of \wobs\ is  not only over one order of magnitude higher than the typical values $\sim0.0002-0.0005$ in the local Universe \citep{Kormendy2013}, but also $\sim5$ times higher than the expected value by the evolutionary trend of $M_{\rm BH}/M_\star$ \citep{McLure2006,Peng2006,Merloni2010,Targett2012,Bongiorno2014,Matsuoka2018}.  We derive the black hole mass growth rate ${\dot{M}}_{\rm BH}$ of \wobs\ from the bolometric luminosity by using the relation $L_{\rm AGN}=(\eta {\dot{M}}_{\rm BH} c^2)/({1-\eta})$ and adopting $\frac{\eta}{1-\eta}=0.1$. We infer ${\dot{M}}_{\rm BH}$ = 49 $M_\odot$ yr$^{-1}$, which suggests a rapid growth of the central SMBH. We then calculate the ratio between the black hole mass growth rate and star formation rate and obtain ${\dot{M}}_{\rm BH}$/SFR = 0.007, which is close to the local $M_{\rm BH}/M_\star$ values. The result indicates that \wobs\ is evolving towards the evolutionary trend of $M_{\rm BH}/M_\star$ (the dashed line in Figure \ref{fig:mbh2mstar}).

\section{Summary and Conclusions}\label{sec:sum}

We present ALMA observations of cold dust and molecular gas and multiwavelength SED analysis in a {\it WISE}-selected, hyperluminous dust-obscured quasar \wobs\ at $z=2.9$. We derive the physical properties of each component, such as molecular gas, stars, dust and the central SMBH. We summarize our main results as follows. 

\begin{enumerate}
\item Our ALMA band-3 observations detect both the dust continuum and the \coline\ line. We derive molecular gas mass, $M_{\rm gas}=8.4\times10^{10}\ M_\odot$ and its fraction $f_{\rm gas}=0.7$ suggest that \wobs\ is gas-rich. Based on the FWHM of the \coline\ line, we estimate the dynamical mass $M_{\rm dyn}=1.6\times10^{10}\ M_\odot$, which is generally consistent with the sum of the molecular gas mass and stellar mass. The velocity map of the \coline\ emission line showing a velocity gradient is possibly due to a rotating gas disk. However, other possibilities, such as mergers and outflows, cannot be ruled out.  Under the assumption of rotation disk, we can roughly estimate the rotation velocity and velocity dispersion. The ratio $V_{\rm rot}/\sigma$ is large, indicating that the gas disk is possibly unstable, which may be partly responsible for the observed high star formation efficiency in \wobs.
\item By assuming a graybody model, we derive the cold dust temperature $T_{\rm dust}=78.1\pm5.6$ K and dust mass $M_{\rm dust}=(8.9\pm1.6)\times10^7\ M_\odot$. The gas-to-dust ratio of \wobs, $\sigma_{\rm GDR}=944\pm77$, is about one order of magnitude higher than the typical values for the Milky Way, the local SFGs/ULIRGs and high-redshift SMGs. 
\item Based on the UV-to-millimeter SED modeling, we derive the stellar mass and SFR of \wobs. The stellar mass of \wobs\ is $(3.5\pm0.9)\times10^{10}\ M_\odot$. The star formation rate (SFR) is estimated to be $\sim3000-7000\ M_\odot$ yr$^{-1}$ by using different methods. The high values of SFR and specific SFR suggest \wobs\ is a maximum-starburst. The corresponding gas depletion timescales are very short ($t_{\rm depl}\sim12-28$ Myr).
\item Finally, we infer the black hole mass growth rate of \wobs\ (${\dot{M}}_{\rm BH}$ = 49 $M_\odot$ yr$^{-1}$), which suggests a rapid growth of the central SMBH. The observed black hole to stellar mass ratio $M_{\rm BH}/M_\star$ of \wobs, which is dependent on the adopted Eddington ratio, is over one order of magnitude higher than the local value if assuming a reasonable Eddington ratio $\lambda_{\rm Edd}=1.0$, and is evolving towards the evolutionary trend of unobscured quasars.
\item Our results suggest that \wobs\ has both a gas-rich maximum-starburst and a rapid black hole growth within it. All results are consistent with the scenario that \wobs\ is experiencing a short transition phase towards an unobscured quasar. 
\end{enumerate}

\acknowledgments

We thank the anonymous referee for constructive comments and suggestions. We thank Dr. Hu Zou (NAOC) for his help on optical photometry. This work is supported by National Key R\&D Program of China (No. 2017YFA0402703). We thank the staff of the Nordic ALMA Regional Center (ARC) node for their support and helpful discussions. The Nordic ARC node is based at Onsala Space Observatory and funded through Swedish Research Council grant No 2017-00648. LF acknowledges the support from the National Natural Science Foundation of China (NSFC, Grant Nos. 11822303, 11773020 and 11433005) and Shandong Provincial Natural Science Foundation, China (ZR2017QA001, JQ201801). KK acknowledges support from the Knut and Alice Wallenberg Foundation and the Swedish Research Council. YH acknowledges the support from NSFC (Grant No. 11773063) and Natural Science Foundation of Yunnan Province (Grant No. 2017FB007). Q.-H.T acknowledges the support from the NSFC (Grant No. 11803090). 

This paper makes use of the following ALMA data: ADS/JAO.ALMA\#2017.1.00441.S. ALMA is a partnership of ESO (representing its member states), NSF (USA) and NINS (Japan), together with NRC (Canada), NSC and ASIAA (Taiwan), and KASI (Republic of Korea), in cooperation with the Republic of Chile. The Joint ALMA Observatory is operated by ESO, AUI/NRAO and NAOJ.

This paper makes use of data products from the Wide-field Infrared Survey Explorer, which is a joint project of the University of California, Los Angeles, and the Jet Propulsion Laboratory/California Institute of Technology, funded by the National Aeronautics and Space Administration.

This paper used public archival data from the Dark Energy Survey (DES). Funding for the DES Projects has been provided by the U.S. Department of Energy, the U.S. National Science Foundation, the Ministry of Science and Education of Spain, the Science and Technology Facilities Council of the United Kingdom, the Higher Education Funding Council for England, the National Center for Supercomputing Applications at the University of Illinois at Urbana-Champaign, the Kavli Institute of Cosmological Physics at the University of Chicago, the Center for Cosmology and Astro-Particle Physics at the Ohio State University, the Mitchell Institute for Fundamental Physics and Astronomy at Texas A\&M University, Financiadora de Estudos e Projetos, Funda{\c c}{\~a}o Carlos Chagas Filho de Amparo {\`a} Pesquisa do Estado do Rio de Janeiro, Conselho Nacional de Desenvolvimento Cient{\'i}fico e Tecnol{\'o}gico and the Minist{\'e}rio da Ci{\^e}ncia, Tecnologia e Inova{\c c}{\~a}o, the Deutsche Forschungsgemeinschaft, and the Collaborating Institutions in the Dark Energy Survey. The Collaborating Institutions are Argonne National Laboratory, the University of California at Santa Cruz, the University of Cambridge, Centro de Investigaciones Energ{\'e}ticas, Medioambientales y Tecnol{\'o}gicas-Madrid, the University of Chicago, University College London, the DES-Brazil Consortium, the University of Edinburgh, the Eidgen{\"o}ssische Technische Hochschule (ETH) Z{\"u}rich,  Fermi National Accelerator Laboratory, the University of Illinois at Urbana-Champaign, the Institut de Ci{\`e}ncies de l'Espai (IEEC/CSIC), the Institut de F{\'i}sica d'Altes Energies, Lawrence Berkeley National Laboratory, the Ludwig-Maximilians Universit{\"a}t M{\"u}nchen and the associated Excellence Cluster Universe, the University of Michigan, the National Optical Astronomy Observatory, the University of Nottingham, The Ohio State University, the OzDES Membership Consortium, the University of Pennsylvania, the University of Portsmouth, SLAC National Accelerator Laboratory, Stanford University, the University of Sussex, and Texas A\&M University. Based in part on observations at Cerro Tololo Inter-American Observatory, National Optical Astronomy Observatory, which is operated by the Association of Universities for Research in Astronomy (AURA) under a cooperative agreement with the National Science Foundation.

%

\vspace{5mm}
\facilities{ALMA, Herschel (PACS, SPIRE), WISE, CTIO (DECam)}

\end{document}